# Disorder-Induced Elevation of Intrinsic Anomalous Hall Conductance in an Electron-Doped Magnetic Weyl Semimetal

Jianlei Shen[1,2†], Qiushi Yao[3†], Qingqi Zeng[1,2], Hongyi Sun[3], Xuekui Xi[1], Guangheng Wu[1], Wenhong Wang[1,4], Qihang Liu[3,5*], Enke Liu[1,4*]

1. State Key Laboratory for Magnetism, Institute of Physics, Chinese Academy of Sciences, Beijing 100190, China

2. University of Chinese Academy of Sciences, Beijing 100049, China

3. Shenzhen Institute for Quantum Science and Technology and Department of Physics, Southern University of Science and Technology (SUSTech), Shenzhen 518055, China

4. Songshan Lake Materials Laboratory, Dongguan, Guangdong 523808, China

5. Guangdong Provincial Key Laboratory for Computational Science and Material Design, Southern University of Science and Technology, Shenzhen 518055, China

[†] These authors contributed equally to this work.
[*] ekliu@iphy.ac.cn (E. L.)
[*] liuqh@sustech.edu.cn (Q. L.)

**Abstract**

Topological materials are expected to show distinct transport signatures due to their unique band-inversion character and band-crossing points. However, the intentional modulation of such topological responses by experimentally feasible means is less explored. Here, an unusual elevation of anomalous Hall effect (AHE) is obtained in electron(Ni)-doped magnetic Weyl semimetal $Co_{3-x}Ni_xSn_2S_2$, showing peak values of anomalous Hall-conductivity, Hall-angle and Hall-factor at a relatively low doping level of $x = 0.11$. The separation of intrinsic and extrinsic contributions to total AHE using TYJ scaling model indicates that such significant enhancement is dominated by the intrinsic mechanism of electronic Berry curvature. Theoretical calculations reveal that compared with the Fermi-level shifting from electron filling, a usually overlooked effect of doping, i.e., local disorder, imposes a striking effect on broadening the bands and narrowing the inverted gap, and thus results in an elevation of the integrated Berry curvature. Our results not only realize the enhancement of AHE in a magnetic Weyl semimetal, but also provide a practical design principle to modulate the bands and transport properties in topological materials, by exploiting the disorder effect of doping.

As a pseudo-magnetic field in the momentum space, the Berry curvature describes the geometric phase of the wavefunction and the topology of the energy bands[1, 2]. In magnetic materials with broken time-reversal symmetry, the integration of the Berry curvature of all the occupied states through the whole Brillouin zone adds a transverse velocity term into the equation of motion, giving rise to intrinsic contribution to the anomalous Hall conductivity (AHC) [2, 3]. In topological materials with an inverted order between conduction and valence bands, the nontrivial Berry curvature is expected to lead to significant contribution to the anomalous Hall effect and the relevant physical properties. Recently, a magnetic Weyl semimetal $Co_3Sn_2S_2$ was experimentally confirmed [4-7], with an giant AHC (~1130 $\Omega^{-1}$ $cm^{-1}$) and anomalous Hall angle (AHA, ~20%). The AHC measured in experiment is highly consistent with the theoretical calculations from the Berry curvature, indicating that the AHC is dominated by intrinsic contribution, which originates from the clean nontrivial bands around the Fermi level ($E_F$) leading to large Berry curvature from the band inversion and Weyl nodes[4].

How to further enhance the transverse transport effect like AHC in magnetic topological materials, especially in $Co_3Sn_2S_2$ that already has a giant value? According to Kubo formula[8], the intrinsic AHC is closely related to two factors. One is the electron occupation determined by the position of $E_F$; the other is the topological band character around $E_F$, such as the size of the inverted band gap. We consider the possibility of doping, which is well established as one of the best ways to control the material properties. In $Co_3Sn_2S_2$, it is predicted that the position of $E_F$ is already optimal for the integrated Berry curvature[4], implying that within the rigid band framework, doping is thus not expected to further elevate the intrinsic AHC. However, compared with the electric voltage gating effect[9, 10], chemical doping-induced modulation of electronic structure is often overlooked. For example, a disordered distribution of the dopants will change the local potential environment and destroy the translational symmetry of the system[11, 12], which inevitably modifies the band structure of the system and affects the intrinsic AHC. Meanwhile, the introduction of alien atoms may cause the impurity scattering, which usually generates an extrinsic contribution to the AHC[13]. Therefore, the physical understanding of doping for the AHC in topological materials is intricated and thus requires comprehensive explorations.

In this Letter, through transport measurements, transport model scaling, and atomistic description of chemical doping theory encoded in first-principles calculations, we find that in $Co_{3-x}Ni_xSn_2S_2$ with Ni atoms substituting Co, the doping behavior simultaneously tunes the electron occupation and the inverted band gap, which give rise to opposite contributions to the AHC. Remarkably, with low doping concentrations ($x \leq 0.1$), disorder-induced band gap narrowing dominates the modulation of Berry curvature, lea、ding to a sizable enhancement of AHC up to 60% compared with the undoped sample. Scaling models elucidate that such enhancement of AHC is mainly from the intrinsic contribution rather than extrinsic one such as skew scattering and side jump. Our work not only provides a practical way to further enhance the AHC and relevant properties by tailoring the Berry curvature in topological semimetals, but also reveals the impact of doping-induced disorder effect to the emergent topological properties.

In order to probe the evolution of AHC with Ni substitution of Co, i.e., $Co_{3-x}Ni_xSn_2S_2$, a series of single crystals of $Co_{3-x}Ni_xSn_2S_2$ ($0 \leq x \leq 0.22$), with an initial mixture of molar ratios of (Co+Ni) : S : Sn : Pb = 12 : 8 : 35 : 45, was grown using Sn and Pb mixed flux. The chemical compositions of these crystals were determined using the energy dispersive X-ray spectroscopy (See Fig. S1 and Table S1). Figure 1(a) shows the rhombohedral structure with the space group of R-3m (No. 166) for $Co_3Sn_2S_2$. The crystal possesses quasi-two-dimensional $Co_3Sn$ layers sandwiched between S atoms, with Ni atoms randomly replacing Co atoms on the kagome lattices. The magnetic moments are mainly located on Co atoms of kagome layer and align along the *c*-axis. The magnetic and electrical properties of crystals were characterized by superconducting quantum interference device (SQUID) magnetometer and the physical property measurement system (PPMS), respectively. Saturation magnetization ($M_S$) and Curie temperature ($T_C$) decrease with increasing Ni content, which is mainly attributed to the weaker spin splitting and exchange interaction of Ni than Co (See Fig. S2). The temperature dependence of longitudinal resistivity ($\rho_{xx}$) for I // *a* and B = 0 shows a rapid drop below the kink point $T_C$, as shown in Fig. 1(b). In addition, the residual electric resistivity increases gradually with increasing Ni content due to the enhancement of dopant scattering.

Figure 1(c) shows the Hall resistivity ($\rho_{yx}$) as a function of magnetic field at 10 K in the configuration of B // *c* and I // *a*. The anomalous Hall resistivities $\rho_{yx}^A$ are then obtained as the zero-field extrapolation of the high field part. With increasing Ni content, the $\rho_{yx}^A$ at 10 K increases clearly from 4.0 μΩ cm of *x* = 0 to 18.2 μΩ cm of *x* = 0.22, as shown in Fig. 1(d). The anomalous Hall conductivity ($\sigma_{yx}^A$) can be further calculated using $\rho_{xx}$ and $\rho_{yx}$ via the relation $\sigma_{yx}^A = -\rho_{yx} / (\rho_{yx}^2 + \rho_{xx}^2)$. The results at 10 K show a first increase and a subsequent decrease with increasing Ni content, forming a maximum of 1382 $\Omega^{-1}$ cm$^{-1}$ at *x* = 0.11 (Fig. 1(d)). The prominent enhancement from 850 to 1382 $\Omega^{-1}$ cm$^{-1}$ shows an increase by more than 60% at the low doping levels. On the other hand, as an important measure of the transport properties of anomalous Hall materials, the AHA reflects the conversion efficiency of longitudinal current to transverse one, which can be characterized by $\sigma_{yx}^A / \sigma_{xx}$. With increasing Ni content, one can see the AHA increases up to 14% at 10 K, showing a small peak at *x* = 0.11, as shown Fig. 1(e). The temperature dependence of AHA can be obtained from the $\rho_{yx}(B)$ data measured at different temperatures (See Fig. S3). For each Ni content, the maximal value of AHA is also shown in Fig. 1(e). Clearly, a sharp peak of AHA appears, with a maximum of 22% also at $x = 0.11$, which is higher than the undoped $Co_3Sn_2S_2$. These experimental results unambiguously show that about 10% Ni doping to topological Weyl semimetal $Co_3Sn_2S_2$ leads to significant enhancement of AHC, indicating that the manipulation of Berry curvature by chemical doping is more complicated than previously thought, i.e., shifting $E_F$ in a rigid band structure.

In general, the anomalous Hall effect can come from both intrinsic effects, i.e., Berry curvature of the ground-state wavefunctions, and extrinsic effects such as skew scattering and side jump[13]. We next apply two methods to decompose the total AHC to intrinsic and extrinsic contributions. For the first one, the relation of $\rho_{yx}^A$ and $\rho_{xx}$ was fitted using the formula $\log \rho_{yx}^A = \alpha \log \rho_{xx}$, where $\alpha$ is exponent of $\rho_{xx}$. For intrinsic mechanism, $\sigma_{yx}^A$ can be scaled by $\sigma_{xx}^2$ [14]. As shown in inset of Fig. 2(a), the value of the scaling exponent $\alpha$ basically keeps at 2 ~ 1.9 for $x = 0$ ~ 0.17, which indicates that the intrinsic contribution originated from Berry curvature dominates the AHC within this doping range. With further increasing Ni content, $\alpha$ drops to 1.63, indicating that the extrinsic contribution from the impurity scattering is enhanced. The evolution of $\alpha$ with Ni doping indicates that while the AHC is dominated by the intrinsic contribution the weak extrinsic contribution gradually increases.

To quantitatively reveal the contributions of intrinsic and extrinsic mechanism, the TYJ scaling model, proposed by Tian et al.[15, 16], was further adopted. As a scaling $\rho_{yx}^A = a\rho_{xx0} + b\rho_{xx}^2$ of the AHE, with $\rho_{xx0}$ being the residual resistivity, TYJ model has been widely used in many systems[17-20]. In this model, the first item on the right of the equation comes from the contribution of the extrinsic mechanism to AHE. The second item comes from the intrinsic contributions of band Berry curvature. The TYJ scaling can also be expressed as $\sigma_{yx}^A = -a\sigma_{xx0}^{-1}\sigma_{xx}^2 - b$, where $\sigma_{xx0} = 1/\rho_{xx0}$ and $b$ is the residual conductivity and intrinsic $\sigma_{yx}^A$, respectively. Figure 2(b) shows the linear relationship between $\sigma_{yx}^A$ and $\sigma_{xx}^2$ for different Ni contents, with the intrinsic $\sigma_{yx}^A$ values obtained by intercept $b$ on the longitudinal axis. As shown in Fig. 2(c), it is obvious that the intrinsic $\sigma_{yx}^A$ ($\sigma_{yx}^A$(int.)) increases first and then decreases with increasing Ni content, also forming a maximum of 1340 $\Omega^{-1}$ cm$^{-1}$ at $x = 0.11$, while further increasing Ni content leads to a decrease of $\sigma_{yx}^A$(int.). One can see both peaks of $\sigma_{yx}^A$(int.) and measured total $\sigma_{yx}^A$ overlap each other in the doping range of $x = 0$ ~ 0.17. Meanwhile, during the Ni doping the extrinsic $\sigma_{yx}^A$ ($\sigma_{yx}^A$(ext.)) basically stays around zero and increases slowly to 214 $\Omega^{-1}$ cm$^{-1}$ at $x = 0.22$. Therefore, both of the scaling models indicate that the AHC of $Co_{3-x}Ni_xSn_2S_2$ is enhanced mainly by the intrinsic contribution from electronic structures, while for the peak values of AHC and AHA around $x = 0.11$ the extrinsic contribution by impurity scattering of Ni dopants is very weak.

Having established that the observed AHC enhancement is caused by the tailoring of the Berry curvature of the ground-state electronic structure, we next applied the atomistic description of chemical doping using state-of-art density functional theory (DFT) approaches to gain the fundamental picture of such doping-induced modulation [21, 22]. Generally, doping by electrons means that the $E_F$ shifts towards the conduction band. However, such definition naturally assumes a rigid band model inactive to the doping process. On the other hand, pushing $E_F$ to a higher energy often leads to reactions of the electronic structure. A famous example is the formation of polarons, where the states of introduced carrier couple with the local distortion of the lattice. Specifically, local distortions inevitably induced by doping can result in the global electronic structure an ensemble averaged property $<P(S_i)>$ from many random configurations $S_i$, rather than a property $P(<S>)$ of averaged structure $<S>$ [23, 24]. Therefore, a comprehensive description of doping, especially for quantities like Berry curvature that is sensitive to the electronic structure, should consider $<P(S_i)>$ that includes disorder effect, local environment and symmetry breaking (even the translational symmetry), etc. Thus, we construct large supercells to capture the experimental conditions after the introduction of dopants (see Supplemental Material for calculation details). Then, we take the "effective band structure" (EBS) method [25-29] to unfold the supercell band structures into the primitive Brillouin zone (BZ) for further theoretical analysis. The advantage of EBS is that it transfers the complex $E$-$k$ dispersion within a small supercell BZ into the spectrum density in the primitive BZ. As we will show below, we find that the doping-induced band modulation is vital to understand the experimentally observed AHC evolution.

The band structure of the pristine $Co_3Sn_2S_2$ is shown in Fig. 3(a). With SOC the nodal lines of $Co_3Sn_2S_2$ caused by band inversion are all gapped except three pairs of isolated Weyl points, in agreement with the previous work[4, 5]. In slightly Ni-doped $Co_3Sn_2S_2$ ($x = 0.056$), extra electron filling moves $E_F$ up macroscopically as expected (Fig. 3(b)). Remarkably, due to the local distortion, a special form of disorder effect induced by doping, Fig. 3(b) also shows significant band splitting along both U–L and L–Γ paths. Such splitting could broaden the band spectra and thus reduce the inverted band gap. More dopants further shift $E_F$ upwards and enhance the band broadening, as shown in Fig. 3(c) (also Fig. S6). Recall that the AHC depends on both of the inverted band gap and the position of $E_F$ (see Supplemental Material), the synergic effect of $E_F$ movements and band gap narrowing determines the evolution of AHC upon doping.

As shown in Fig. 3(d), the peak value of the energy dependent AHC of pristine $Co_3Sn_2S_2$ calculated by integrating the Berry curvature of all the occupied bands is exactly located at $E_F$. If we resort to the rigid band model[30], the Co substitution by Ni introduces extra electrons and thus moves $E_F$ to a higher energy, the AHC should decrease accordingly, as mentioned before. To confirm this, we also applied the virtual crystal approximation (VCA) approach[31, 32], which considers a symmetry-preserved primitive cell composed by "virtual" atoms that interpolate between the behavior of the atoms in the parent systems, to simulate the doping effect. Compared with the supercell approach, VCA method counts the electron occupation correctly, but excludes all the doping-induced disorder effects. Indeed, we found that the band structure of doped $Co_3Sn_2S_2$ only captures the feature of $E_F$ movements within an almost intact band structure (see Fig. S4). As a result, the calculated AHC decreases monotonically with increasing Ni doping (see Fig. S5), which is inconsistent with the experimental observation. In sharp contrast, our supercell approach predicted an AHC enhancement at $x = 0.056$, indicating that the doping-induced modulation of the electronic structure is indeed important to enhance the Berry curvature. Furthermore, the supercell approach also reproduces the AHC evolution with a maximum compared with that of the intrinsic component decomposed from the measured AHC except for a slightly shifted peak position, as shown in Figs. 3(d) and 3(e).

Overall, our supercell approach captures both the $E_F$-shifting effect from extra electron filling and the band broadening effect from disorder and local distortions. The former (latter) factor tends to reduce (enhance) the intrinsic contribution to AHC of $Co_3Sn_2S_2$, leading to a competing mechanism of the evolution of intrinsic AHC, as illustrated in Fig. 3(f). At low doping levels, as long as that $E_F$ is still located mostly within the inverted gap, the total AHC is dominated by the gap narrowing from band broadening, leading to an increasing trend. This can be confirmed by the fact that the peak of energy-dependent AHC for $x = 0.056$ is also located at $E_F$ (see Fig. 3(d)). On the other hand, at high doping levels $E_F$ moves away from the inverted gap region and thus strongly suppresses the total AHC despite of an even smaller inverted band gap.

From the conventional viewpoint[13], the anomalous Hall effect is proportional to $M$ according to $\rho_{yx} = \rho_{yx}^N + \rho_{yx}^A = R_0 B + 4\pi R_S M$, where $R_0$, $B$, $R_S$ and $M$ denote ordinary Hall coefficient, magnetic field, anomalous Hall coefficient and magnetization, respectively. However, as the first experimentally confirmed magnetic Weyl semimetal, $Co_3Sn_2S_2$ exhibits giant anomalous Hall effect[4] and anomalous Nernst effect[33] owing to the nontrivial Berry curvature of the inverted band structure. As a result, the giant $\sigma_{xy}^A$ (1130 $\Omega^{-1}$ cm$^{-1}$) in pristine $Co_3Sn_2S_2$ is accompanied by a small $M$ of only 0.92 $\mu_B$/f.u., lower than many ferromagnetic materials. The anomalous Hall factor $S_H$, defined by $\sigma_{xy}^A$ per unit magnetic moment ($\sigma_{xy}^A / M$), is as high as 1.1 V$^{-1}$, 1 ~ 2 orders in magnitude higher than many other anomalous Hall materials. As shown in Fig. 4, both the values of $S_H$ and AHA in $Co_3Sn_2S_2$ are seen to be the largest by a prominent margin. By tailoring the Berry curvature of topological bands through Ni doping, the giant AHA of $Co_3Sn_2S_2$ are further enhanced to 22% at $x$ = 0.11, accompanied by a decreasing $M$ due to the lower magnetic moments of Ni than Co atoms (see Table S2). Meanwhile, $S_H$ is also bolstered to a maximal value of 2.3 $V^{-1}$ at $x$ = 0.11, as twice large as that of the pristine $Co_3Sn_2S_2$. Such a large enhancement of $S_H$ is attributed to both the reduced magnetization due to the lower-moment Ni doping and the increased AHC owing to the disorder-enhanced Berry curvature from the topological band structure.

In summary, in Ni-doped magnetic Weyl semimetal $Co_3Sn_2S_2$, enhancements of anomalous Hall conductivity ($AHC_{max}$ ~ 1400 $\Omega^{-1}$ $cm^{-1}$), anomalous Hall angle ($AHA_{max}$ ~ 22%) and anomalous Hall factor ($S_{Hmax}$ ~ 2.3 $V^{-1}$) were achieved by the doping-induced modulation of topological band structures. The separation of intrinsic and extrinsic contributions to AHC based on TJY scaling model indicates that the enhancement of AHE is mainly attributed to the intrinsic contribution. Theoretical calculations based on supercell doping approach uncover that the increase of intrinsic contribution, dictated by Berry curvature, originates from the broadening of the inverted bands caused by disorder effects. Such a mechanism, together with the extra electron filling that shifts $E_F$, leads to a collaborative modulation of the anomalous Hall effects observed in Ni-doped $Co_3Sn_2S_2$. Our findings provide a novel understanding on the disorder effect of chemical doping to the modulation of topological bands, which also sheds light on relevant physical properties such as anomalous Nernst effect and magneto-optical Kerr effect in emerging topological magnets.

**Acknowledgements**

This work was supported by National Natural Science Foundation of China (No. 11974394), National Key R&D Program of China (Nos. 2019YFA0704904 and 2017YFA0206303), Beijing Natural Science Foundation (No. Z190009), Users with Excellence Program of Hefei Science Center CAS (No. 2019HSC-UE009), Youth Innovation Promotion Association of CAS (No. 2013002), and Fujian Institute of Innovation, CAS. Work at SUSTech was supported by Guangdong Innovative and Entrepreneurial Research Team Program (No. 2017ZT07C062), the Guangdong Provincial Key Laboratory of Computational Science and Material Design (No. 2019B030301001) and Center for Computational Science and Engineering of Southern University of Science and Technology.

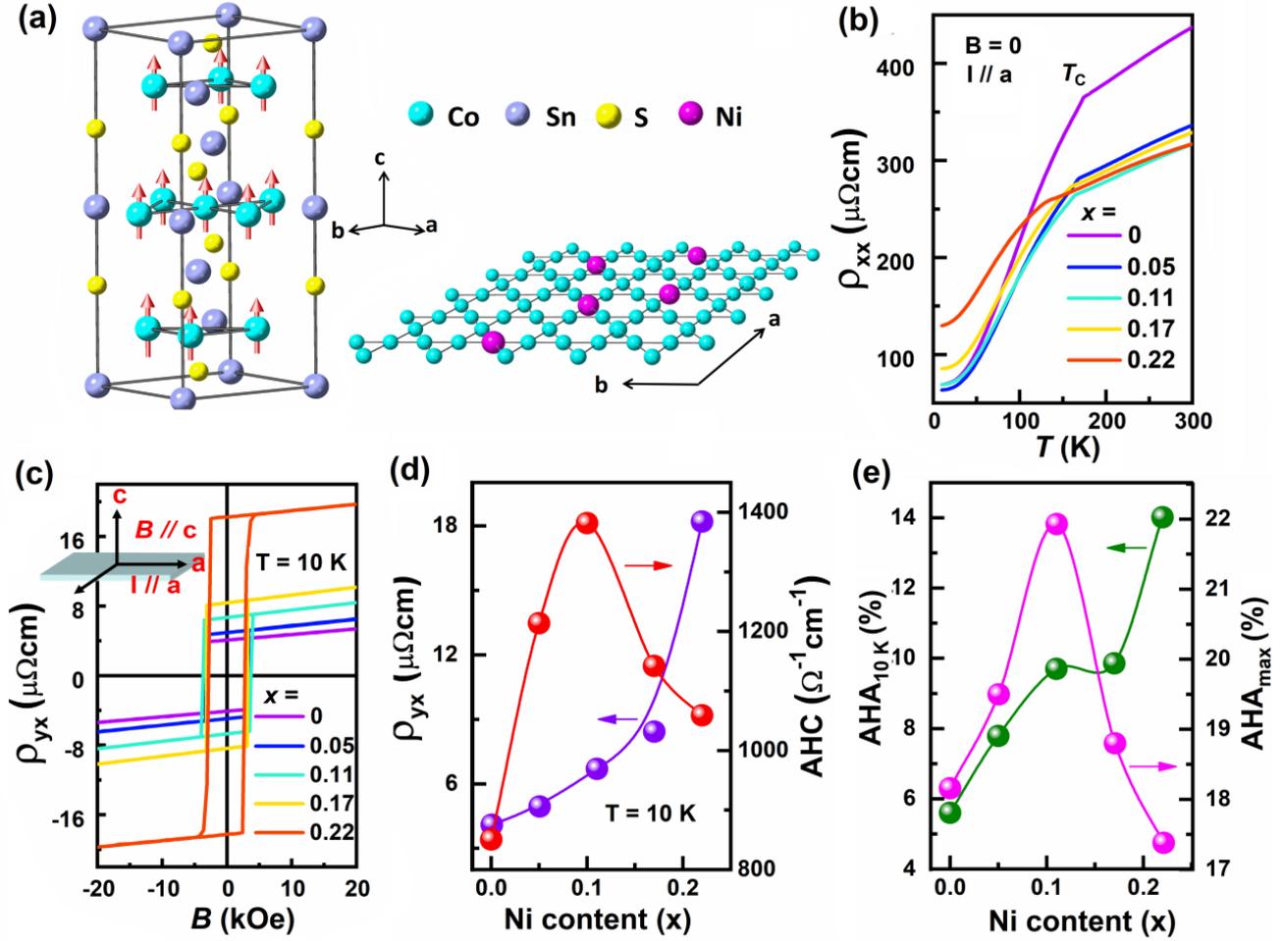

FIG. 1. (a) Crystal structure of $Co_3Sn_2S_2$ and kagome layer composed of Co/Ni atoms. (b) Temperature dependence of $\rho_{xx}$ for different Ni contents. (c) Magnetic field dependence of $\rho_{xy}$ for $B \mathbin{/\mkern-5mu/} c$ and $I \mathbin{/\mkern-5mu/} a$. (d) Ni content dependence of $\rho_{xy}$ and $\sigma_{yx}^A$. (e) Ni content dependence of AHA at 10 K and maximal values of AHA.

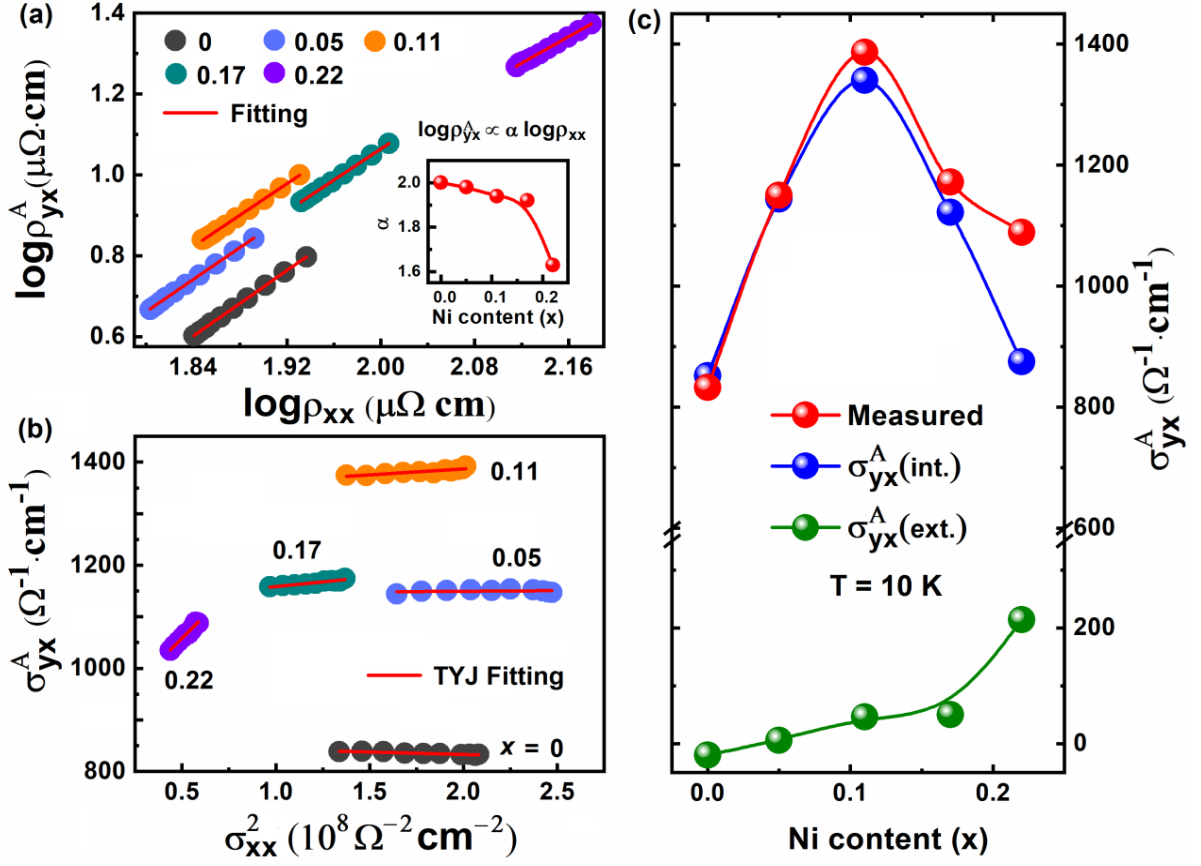

FIG. 2. (a) Plot of log $\rho_{xx}$ vs log $\rho_{yx}^A$. The red solid line is the fitting by relation log $\rho_{yx}^A = \alpha$ log $\rho_{xx}$. Inset shows Ni content dependence of $\alpha$. (b) $\sigma_{yx}^A$ as a function of $\sigma_{xx}^2$ and the fitting by TYJ model. (c) Ni content dependences of intrinsic and extrinsic AHCs, separated by TYJ model from the measured total AHC at 10 K.

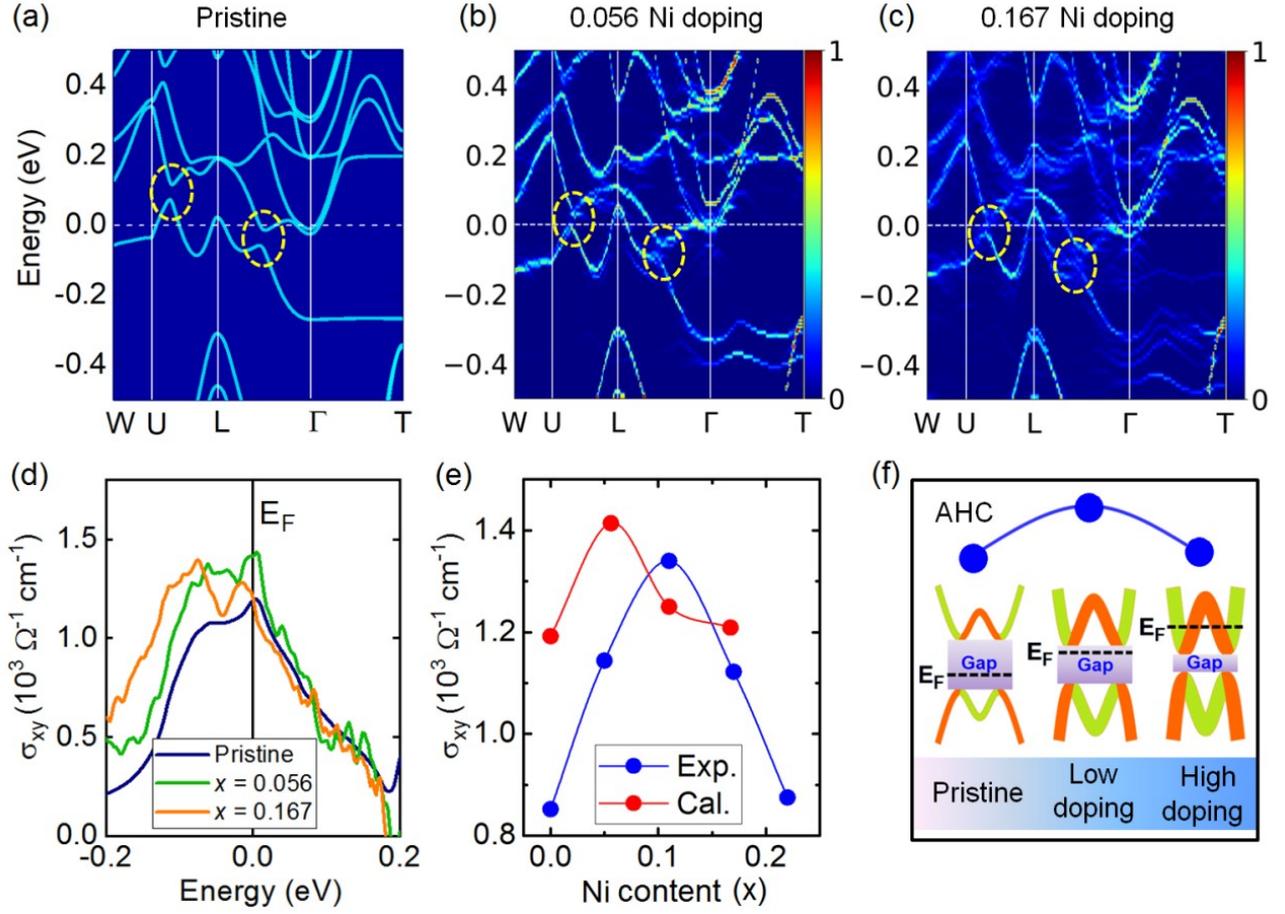

FIG. 3. (a) Band structures of the pristine $Co_3Sn_2S_2$. (b-c) Effective band structure for $x = 0.056$ and $0.167$ Ni-doped $Co_3Sn_2S_2$, respectively. (d) Energy dependent AHC of the pristine, $x = 0.056$ and $0.167$ Ni-doped $Co_3Sn_2S_2$. (e) Evolution of the AHC as a function of doping content from calculation and experiment. (f) Schematic plot showing the modulation effects of band structure and the resultant AHC upon Ni doping. SOC is included for all calculations.

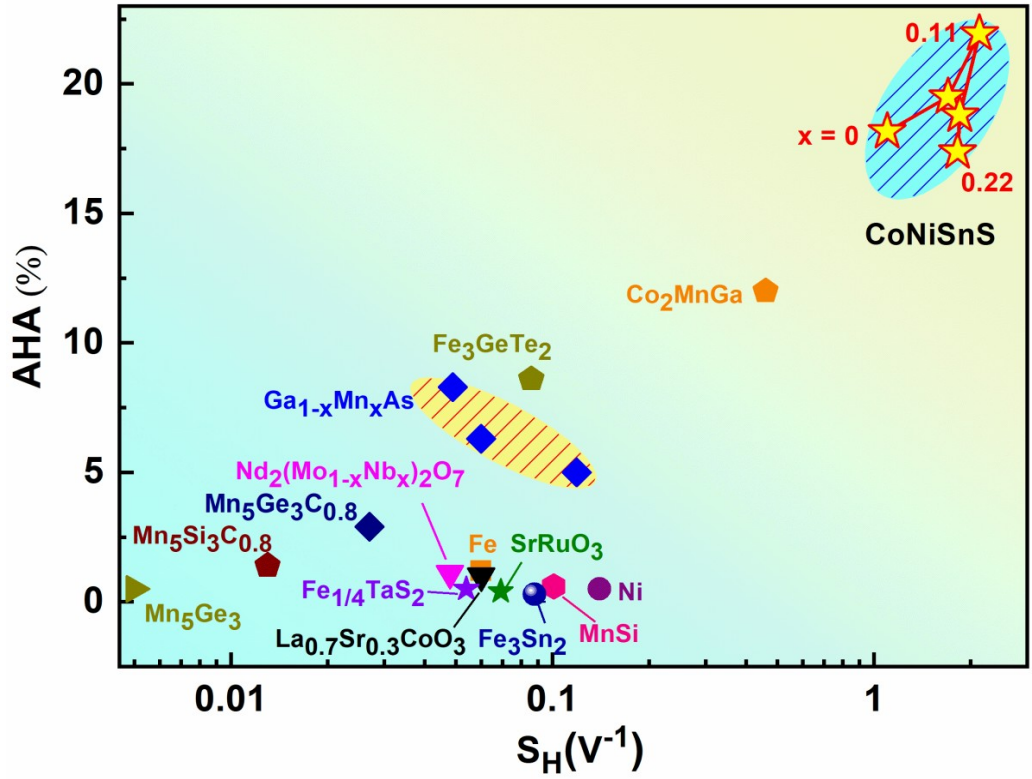

FIG. 4. Comparison of the anomalous Hall angle (AHA) and anomalous Hall factor $S_H$ among $Co_{3-x}Ni_xSn_2S_2$ and other magnetic systems. The reported data were taken from references that can be found in Table S3.